\documentclass[12pt]{article}

\usepackage{amsmath}
\usepackage{mathptmx}
\usepackage{helvet}
\usepackage{courier}
\usepackage{amssymb}
\usepackage{pgfplots}
\usepackage{amsthm}
\usepackage[T1]{fontenc}
\usepackage[latin9]{inputenc}

\usepackage{makeidx}         
\usepackage{graphicx}        
\usepackage{multicol}        
\usepackage[bottom]{footmisc}

\usepackage{caption}
\usepackage{subcaption}

\newtheorem{definition}{Definition}
\newtheorem{theorem}{Theorem}
\newtheorem{remark}{Remark}
\newtheorem{lemma}{Lemma}
\usepackage{hyperref}
\hypersetup{colorlinks=true,
	linkcolor=blue,
	anchorcolor=blue,
	citecolor=red,
	urlcolor=brown,
	pdfauthor={Adriano Barra}}

\providecommand{\be}{\begin{equation}}
  \providecommand{\ee}{\end{equation}}
\providecommand{\bea}{\begin{eqnarray}}
  \providecommand{\eea}{\end{eqnarray}}
\providecommand{\beas}{\begin{eqnarray*}}
  \providecommand{\eeas}{\end{eqnarray*}}

\providecommand{\beni}{\begin{equation*}}
  \providecommand{\eeni}{\end{equation*}}

\providecommand{\bw}{\begin{widetext}}
  \providecommand{\ew}{\end{widetext}}

\newtheorem{Proposition}{Proposition}

\makeindex             

\begin{document}

\author{Adriano Barra\footnote{Dipartimento di Fisica, Sapienza Universit\`{a} di Roma, Roma, Italy and GNFM Gruppo di Roma2.}, Andrea Di Lorenzo\footnote{Dipartimento di Matematica, Sapienza Universit\`{a} di Roma, Roma, Italy.}, \\ Francesco Guerra\footnote{Dipartimento di Fisica, Sapienza Universit\`{a} di Roma, Roma, Italy and INFN Sezione di Roma.}, Antonio Moro\footnote{Department of Mathematics and Information Sciences, University of Northumbria Newcastle, UK}}

\title{On quantum and relativistic mechanical analogues in mean field spin models\footnote{{\em \large The Authors are pleased to dedicate this work to Sandro Graffi in honor of his seventieth birthday}.}.}
\maketitle






\begin{abstract}
Conceptual analogies among statistical mechanics and classical or quantum mechanics often appeared in the literature.
For classical two-body mean field models, such an analogy is based on the identification between the free energy of Curie-Weiss type magnetic models and  the Hamilton-Jacobi action for a one dimensional mechanical system. Similarly, the partition function plays the role of the wave function in quantum mechanics and satisfies the heat equation that plays, in this context, the role of the Schr\"odinger equation.

We show that this identification can be remarkably extended to include a wider family of magnetic models that are classified by normal forms of suitable real algebraic {\it dispersion curves}. In all these cases, the model turns out to be completely solvable as the free energy as well as the order parameter are obtained as solutions of an integrable nonlinear PDE of Hamilton-Jacobi type. We observe that the mechanical analog of these models can be viewed as the {\it relativistic} analog of the Curie-Weiss model and this helps to clarify the connection between generalized self-averaging in statistical thermodynamics and the semiclassical dynamics of viscous conservation laws.
\end{abstract}

\section{Introduction}


A powerful approach for mean field spin glass models is based on formal analogy between mean-field statistical mechanics and the Hamilton-Jacobi formulation of classical mechanics.
\newline
Such an analogy has been pointed out and investigated over a few decades and tracing back in time the genesis of such an approach, due to the vast popularity of these magnetic mean field models, might be not a simple task. Brankov and Zagrebnov in $1983$ used the analogy to accurately describe the Husimi-Temperley model in \cite{bogo3}\footnote{Husimi-Temperley model is the mean field ferromagnet most known as Curie-Weiss model.} and also Newman already pointed out this analogy early in the eighties and even more recently Choquart and Wagner in $2004$ \cite{barbone} and the present authors and colleagues (see~\cite{Gsumrules,barraJSP,genovese,Fourier,aldo} and also~\cite{moro1,moro2,moro3} and~\cite{shannon}).
\newline
However, the discovery of such an analogy turns out to be nothing but the tip of an iceberg demanding a further exploration. This correspondence is indeed very profound and shows a hidden (and at a first glance even counter-intuitive) relation between the Minimum Action Principle in Mechanics (that is often used to describe determinism) and the Second Principle of Thermodynamics (which is often used to justify randomness and stochasticity). Indeed, one can show that, the free energy of a statistical mechanical model can be interpreted as the Hamilton-Jacobi function of a suitable one dimensional mechanical system. For the Curie-Weiss model the Hamilton-Jacobi equations imply that the magnetization satisfies the celebrated Burgers equation, perhaps the simplest scalar model for the propagation of nonlinear waves in a viscosity regime. The thermodynamic limit for  the magnetic model is equivalent to the inviscid limit of the Burgers equation and leads to the so-called inviscid Burgers equation that is also known as the Riemann-Hopf equation. This limit is interpreted as a Second Principle prescription as it turns out to be equivalent to a minimal action principle for the free energy functional. The Riemann-Hopf equation is the simplest example of nonlinear conservation law introduced to describe the propagation of nonlinear hyperbolic waves in the zero dispersion regime. Despite its simplicity, this equation possesses already several interesting features that make it suitable for the description of thermodynamic phase transitions. For instance, solutions to the Rieman-Hopf equation generically break as they develop a gradient catastrophe in finite time. The gradient catastrophe point is associated to caustics of the characteristic lines and it is naturally interpreted as the critical point for a magnetic phase transition. The critical point develops into a classical shock wave that explains the mechanism responsible for discontinuities of the order parameter or its derivatives.

 A model based on the Riemann-Hopf equation is completely integrable via the characteristics method and its general solution provides the equation of state, that is the consistency equation, of the model. This descriptions seems to be very general, as it has also been observed in the context of van der Waals models and its virial extensions~\cite{moro2}  and in pure glassy scenarios~\cite{aldo} and leads to the construction of a one to one correspondence table between some standard concepts in classical thermodynamics and the theory of classical shocks and conservation laws \cite{moro1}. Although the Riemann-Hopf equation turns out to provide an accurate description of the model away from the critical region, in the vicinity of the critical point a suitable multi-scale asymptotic analysis of the Burgers equation is required. It was shown in~\cite{I} that the asymptotic behavior in the vicinity of the critical point is universally expressed in terms of the Pearcey integral and it is argued in \cite{DE} (see~\cite{ALM} too) that such description extends to more general Burgers type equations.


In the present paper, we work out the formal analogy between mean field models and one dimensional mechanical systems at the level of the partition function that in this context plays the role of a (real-valued) quantum-mechanical wave function and satisfies a linear PDE. Consistently with the description outlined above, the associated Hamilton-Jacobi function is interpreted a the free energy of the model. In particular, we focus on a class of solvable generalized models of $N$ interacting spins where the Hamiltonian function is given, as in the cases mentioned above, by the linear combination of the potential associated to the internal spin interaction and the one associated to the external field
\begin{equation}
H_{N} = H_{\textup{int}}(m_{N}) + h H_{\textup{ext}}(m_{N}),
\end{equation}
where
\[
m_{N} = \frac{\sum_{i} \sigma_{i}}{N}
\]
is the mean magnetization per spin particle. We argue that a natural generalization of the Curie-Weiss model can be obtained by the request that the internal and external potentials satisfy a certain polynomial relation referred to as {\it dispersion curve}. This implies, as for the Curie-Weiss model, that the partition  function solves a linear PDE,    where temperature and external magnetic field coupling are the independent variables and the number of particles $N$ plays the role of a scale parameter. The solution in the large $N$ limit is obtained via the standard WKB approach leading to a Hamilton-Jacobi type equation for the free energy function. Similarly to the semiclassical approximation of quantum mechanical models and the geometric optics approximation of the Maxwell equations, the Hamilton-Jacobi type equation so obtained provides an accurate description of the magnetic system in the thermodynamic limit away from the caustic lines associated with the boundary of the critical region.
We analyze in detail models associated to a second order dispersion curve whose normal form reduces to a conic. We note that the parabolic case, referred to as $F-type$ scenario gives the Curie-Weiss model. The elliptic and the hyperbolic case, $P-type$ and $K-type$ scenario respectively (i.e. Poisson-like and Klein-Gordon-like), can be viewed as a deformation of the Curie-Weiss model involving infinitely many $p-$spin contributions. We observe that in all cases the Hamilton-Jacobi type equation for the free energy reduces to a Riemann-Hopf type equation for the expected value of the magnetization. The model is then completely integrable via the characteristics method (see e.g.,~\cite{Whi}) and the critical point of gradient catastrophe is the signature of the occurrence of a magnetic phase transition.

The paper is structured as follows: In Section $2$ we illustrate the methodology in general terms. Section $3$ is dedicated to examples, one for each case. Section $4$ contains our conclusions and outlooks.

\section{Generalized models and techniques for mean field many-body problems}
Given $N$ Ising spins $\sigma_i = \pm 1$, $i \in \{1,...,N\}$, let us consider a general {\em ferromagnetic} model of Hamiltonian of the form
\begin{equation}
\label{Ham}
\frac{H_{N}}{N} = - F(m_{N}) - h G(m_{N}),
\end{equation}
where
\[
m_{N} = \frac{1}{N}\sum_{i=1}^{N} \sigma_{i}
\]
is the magnetization, $F(m_{N})$ models the generic $p-$spin mean field interaction, $G(m_{N})$ accounts for
the interaction with an external magnetic field $h$ (that in many cases is one-body, i.e., $G(m_N) = m_N$).

Note that, generally, with the adjective {\em ferromagnetic} we mean models whose interaction matrix has only positive entries, e.g., $H_N = -(1/N) \sum_{i<j}^N J_{ij} \sigma_i \sigma_j$, with $J_{ij}= J>0$ for all the $N(N-1)/2$ couples. However, as the effect of $J$ on the model's thermodynamics is only to shift the critical temperature, in the following we simply set $J \equiv 1$.

The Boltzmann average of the magnetization is standardly denoted as follows
\be
\langle m \rangle = \lim_{N \to \infty}\frac{\sum_{\{ \sigma \}}^{2^N}\sigma_i \exp(-\beta H_N)}{\sum_{\{ \sigma \}}^{2^N} \exp(-\beta H_N)},
\ee
where the sum is evaluated over all spin configurations $\{\sigma\}$, and $\beta =1/k_B T$ where $T$ is the temperature and $k_B$ is the Boltzmann constant (that we set to one in proper units).
The main object of interest is the free energy function $f(\beta,h)= -\alpha(\beta,h)/\beta$, where
\be
\label{free_en}
\alpha(\beta,h)=\lim_{N \to \infty} \frac1N \ln \sum_{\{ \sigma \}}^{2^N}\exp(-\beta H_N),
\ee
is called {\em mathematical pressure}.
\newline
The free energy is related to the thermodynamical averages of intensive entropy $S$ and internal energy $E$ via the standard formula$f= E - \beta^{-1} S$ (or, alternatively in terms of the mathematical pressure,  $\alpha(\beta, h)= S - \beta E$) that allows to deduce all thermodynamic properties of the system induced by the Hamiltonian $H_{N}$. However, as the mathematical pressure $\alpha(\beta, h)$ is more convenient for computational purposes w.r.t. $f(\beta,h)$, and its usage largely prevailed in the community of disordered statistical mechanics (where most of the applications -of the theory we are going to develop- lie) in the following we will use the former with a little language abuse.

\subsection{Generalized thermodynamic limit and its variational formulation}

Once introduced two scalar variables $t \in \mathcal{R}^+$ and $x \in \mathcal{R}$ (which can be though as time and space in the {\em mechanical analogy} that we are going to develop), we consider at first $F$ such that $F(m)=F(-m)$, $\partial^{2}_{xx}F(m)>0$ and $F(0)=0$, and we set  $G(m)\equiv m$; then we consider the class of models associated to an Hamiltonian $-N[F(m)+h m]\equiv H \colon (0,1) \ni m \mapsto H(m)$.
\newline
We now prove that under the above assumptions the thermodynamic limit for the system defined via $H_{N}$ is well defined. We have the  following
 \begin{theorem}
\label{THtherm}
The thermodynamic limit for the free energy $\alpha_N(t,x)$ exists and reads as
\be
\lim_{N \to \infty} \frac1N \ln Z_N(t,x) = \inf_{N} \frac1N \ln Z_N(t,x) = \alpha(t,x).
\ee
where $Z(x,t)$ is the partition function
\be
\label{Zdef}
Z_N(t,x) = \sum_{\{ \sigma \}}^{2^N}\exp\left(N ( t F(m_{N}) + x m_{N}  ) \right),
\ee
defined $\forall t >0$ and $\forall x \in \mathcal{R}$ (that, in order to bridge with thermodynamics should be related to temperature and magnetic field via $t = 1/T$ and $x = h/T$).
\end{theorem}

The proof of this statement works within the classical Guerra-Toninelli scheme \cite{hightempGT}.
It is sufficient  to prove the model sub-additivity as stated in the following
\begin{lemma}\label{lemma1}
The extensive free energy related to the generalized models defined by $-H(m)/N=F(m)+hG(m)$ is sub-additive in the volume $N$, namely
\be
\ln Z_N(t,x) \leq \ln Z_{N_1}(t,x) + \ln Z_{N_2}(t,x).
\ee
\end{lemma}
\proof
Let us split the system in two subsystems of size $N_1$ and $N_2$ such that $N=N_1+N_2$. Let us $m_1$ and $m_{2}$ be the partial magnetizations associated to the two subsystems such that $m = (N_1/N) m_1 + (N_2/N) m_2,$. Hence, due to convexity of $F$, we have
\be
\label{Fineq}
F(m)=F \left (\frac{N_1}{N} m_1 + \frac{N_2}{N} m_2 \right) \leq \frac{N_1}{N} F(m_1) + \frac{N_2}{N} F(m_2).
\ee
In virtue of the above inequality, the partition function~(\ref{Zdef}) satisfies the following
\be
Z_N(t,x) \leq Z_{N_1}(t,x) \cdot Z_{N_2}(t,x),
\ee
that proves the lemma. \endproof
The route from lemma \ref{lemma1} to theorem \ref{THtherm} is the classical one paved by Ruelle \cite{ruelle}.

Now we proceed showing that the variational formulation of statistical mechanics is
preserved even in this extended scenario. Let us prove the following
\begin{theorem}
Given the variational parameter $-1 \leq M \leq +1$ and the trial free energy
\be
\tilde{\alpha}(t,x|M) = \ln 2 + \ln\cosh \left(x + t \partial_x F(M)\right)+t\left(F(M)-M \partial_x F(M)\right),
\ee
and its optimized value (w.r.t. $M$) as
$$\hat{\alpha}(t,x) = \max_{M}\tilde{\alpha}(t,x|M),$$
then we can write $\alpha(t,x)=\hat{\alpha}(t,x) $.
\end{theorem}
\proof
Let us introduce the auxiliary function $g(m,M)$ as
\be
g(m,M) = \exp\left( -t N \left(F(m)-F(M)  - \partial_x F(M) \left(m-M\right)\right) \right).
\ee
Clearly, due to convexity we have $g(m,M) \leq 1$. Let us consider only those values of $M$ that can also be assumed by $m$ and  let us restrict only on those values the sum over $M$, which will be denoted with a star, i.e. so $\sum_M \to \sum_M^*$. Then
\be\label{over}
\sum_M^* g(m,M) \geq 1,
\ee
because, with probability one, a term in the sum will have $m=M$ and its corresponding $g(M,M)\equiv 1$, as all the others are non-negative, eq.~(\ref{over}) holds. Then, we have
\be
Z_N(t,x) =\sum_{\sigma} e^{t N F(m)} e^{x N m}\geq \sum_{\sigma} e^{t N F(m)} e^{x N m}g(m,M) = e^{N \tilde{\alpha}_N(t,x|M)},
\ee
as $1 \geq g(m,M)$, thus the sum factorizes, $F(m)$ terms cancel and we can conclude the first bound, namely, taking the thermodynamic limit and optimizing w.r.t. $M$
\be
\alpha(t,x) \geq \hat{\alpha}(t,x).
\ee
To prove the reverse bound we can write
\be
Z_N(t,x) \leq \sum_{\sigma}e^{t N F(m)} e^{x N m} \sum_M^* g(m,M) = \sum_M^* e^{N \hat{\alpha}(t,x|M)}\leq \sum_M^* e^{N \hat{\alpha}(t,x)},
\ee
thus $\alpha_N(t,x) \leq \hat{\alpha} + \ln(1+N)/N$ because the $\sum_M^*$ now gives $N+1$ identical terms (as in the last passage there is no longer dependence by $M$ during the summation procedure), hence $Z_N(t,x) \le (N+1) \exp {N\hat{\alpha}(t,x)}$: taking the logarithm of $Z_N(t,x)$ and dividing by $N$, we obtain the expression above, which in the thermodynamic limit returns the expected bound and closes the proof.
\endproof
The study of those values of $M(t,x)$ that optimize the evolution will then be achieved in the following subsections through the mechanical approach.

\subsection{Dispersion curve and generalized models}

 Let us assume that the potentials $F(m_{N})$ and $G(m_{N})$ that define the Hamiltonian~(\ref{Ham}) belong to the {\it dispersion curve}  given by the equation
\begin{equation}
\label{Pdcond}
P_{d}(F,G) = 0
\end{equation}
where
\[
P_{d}(\eta, \xi) = \sum_{k,l} c_{k,l} \eta^{k} \xi^{l}
\]
is a polynomial of degree $d = \max \{k+l \; | \; c_{k,l} \neq 0\}$. Introducing the linear differential operator of order $d$
\[
L_{d} = \sum_{k,l} c_{k,l}  \partial_{t}^{k} (-\partial_{x})^{l},
\]
one can readily verify that, given the condition (\ref{Pdcond}), the partition function~(\ref{Zdef})  can be obtained as a solution to the following linear differential equation
\begin{equation}
\label{Leq}
L_{d} \left [Z_{N} \right] = 0.
\end{equation}
The equation~(\ref{Leq}) can be viewed as the statistical analog of a quantum mechanical wave equation where $Z_{N}$ plays the role of the wave function. More explicitly, setting $\nu = 1/N$, the equation~(\ref{Leq}) reads as follows
\begin{equation}
\label{Leq2}
\sum_{k,l} \nu^{k+l} c_{k,l} \partial_{t}^{k} (-\partial_{x})^{l} Z_{N} = 0.
\end{equation}
From the definition of the free energy $\alpha_{N}$ in~(\ref{free_en}) we get $\alpha_{N} = \nu \log Z_{N}$ and then $Z_{N} = e^{\alpha_{N}/\nu}$.

Substituting the above change of variable into eq.~(\ref{Leq2}), we obtain at the leading order as $\nu \to 0$ (according to the standard WKB approximation) the following Hamilton-Jacobi type equation
\[
P_{d} \left(\alpha_{t}, \alpha_{x} \right) = 0,
\]
where $\alpha = \lim_{N \to \infty} \alpha_{N}$.

Let us now analyze the particular class of models associated to a polynomial relation of the form~(\ref{Pdcond}) of degree $d=2$, that is
\begin{equation}
\label{poly2}
c_{1} F^{2} + c_{2} F G + c_{3} G^{2} + c_{4} F + c_{5} G + c_{6} = 0.
\end{equation}
The quadratic equation~(\ref{poly2}) can be reduced via a suitable linear change of variables  to one of the following canonical forms
\begin{align}
&F^{2} + G^{2} - 1 =0\\
&F^{2} - G^{2} - 1 =0\\
&F -G^{2} = 0.
\end{align}
The corresponding partition function satisfies one the following normal forms
\begin{subequations}
\label{PKFtype}
\begin{align}
\label{Ptype}
\nu^{2} \left( Z_{tt} + Z_{xx} \right ) &= Z,\\
\label{Ktype}
\nu^{2} \left( Z_{tt} - Z_{xx} \right) &= Z,\\
\label{Ftype}
Z_{t} - \nu Z_{xx} &= 0.
\end{align}
\end{subequations}
Many body problems associated to a quadratic dispersive curve will be referred to as $P-$type, $K-$type and $F-$type according to whether their canonical form is the Poisson equation~(\ref{Ptype}), the Klein-Gordon equation~(\ref{Ktype}) and the Fourier (or heat) equation~(\ref{Ftype}) respectively.
\begin{Proposition}
The WKB approximation of equations~(\ref{PKFtype}), standardly performed by the substitution
$Z = e^{\alpha/\nu}$
gives, in the thermodynamic limit $\nu \to 0$ (i.e. $N \to \infty$), one of the following three equations for the free energy $\alpha$

\begin{subequations}
\label{PKFlead}
\begin{align}
\label{Plead}
\alpha_{t}^{2} + \alpha_{x}^{2} &=1, \\
\label{Klead}
\alpha_{t}^{2} - \alpha_{x}^{2} &=1, \\\
\label{Flead}
\alpha_{t} - \alpha_{x}^2 &=0.
\end{align}
\end{subequations}

\end{Proposition}
Equations~(\ref{PKFlead}) show that the free energy $\alpha$ plays the same role as the Hamilton-Jacobi function in classical mechanics.

Moreover, equations~(\ref{PKFlead}) are completely integrable and can be solved via the method of characteristics. Differentiating equations~(\ref{PKFlead}) w.r.t. $x$, we obtain the following Riemann-Hopf type equation
\begin{equation}
\label{PKFhopf}
u_{t} = \left(V(u) \right)_{x}
\end{equation}
where $u = \alpha_{x}$ and the function $V(u)$ is given as follows
\begin{align*}
&\textup{P-type} \ \qquad V(u) =-\sqrt{1-u^{2}} \\
&\textup{K-type} \qquad V(u) =\sqrt{1 + u^{2}} \\
&\textup{F-type} \ \qquad V(u) = u^{2}.
\end{align*}
In particular, based on the classical method of characteristics we have the following
\begin{theorem}
The general solution $u$ to the equation~(\ref{PKFhopf}) is readily obtained via the method of characteristics and it is given by the formula
\begin{equation}
\label{hod}
x + V'(u) t = f(u),
\end{equation}
where $f(u)$ is an arbitrary function of its argument that is locally fixed by the initial condition on $u$. In particular,
given the initial datum
\[
u(x,0) = U(x),
\]
we have that $f = U^{-1}$ is the inverse function of $U(x)$. The free energy, solution to the corresponding equation in~(\ref{PKFlead}),  is obtained by direct integration as follows
\[
\alpha =  \int_0^{x} u(\xi,t) \; d \xi + \Phi(t),
\]
where the the function $\Phi(t)$ is such that $\Phi' = V(u(0,t))$.
\end{theorem}
It is well known that the generic solution to the conservation laws of the form~(\ref{PKFhopf}) breaks in finite time by developing a gradient catastrophe. At the point of the gradient catastrophe that is the analogue of caustics in the Geometric Optics limit and in the semiclassical limit of Quantum Mechanics, the WKB approximation fails and the classical solution develops a multi-valuedness. The appropriate description of the system beyond  the region where the classical solution is multi-valued requires the study of equations~(\ref{PKFtype}). However, the critical point of gradient catastrophe is the signature of a phase transition from a disordered  (``classical") to an ordered (``quantum") state. Clearly, whether or not the a phase transition will occur depends on the particular model that is specified by the initial datum via the function $f(u)$ in~(\ref{hod}). More speifically, we have the following
\begin{theorem}
The critical point $(x_{c}, t_{c}, u_{c})$ is given, if it exists, is a solution to the following equations
\begin{equation}
\label{hodcrit}
x_{c} + V'(u_{c}) t_{c} =f(u_{c}), \qquad V''(u_{c}) t = f'(u_{c}), \qquad V'''(u_{c}) t = f''(u_{c})
\end{equation}
such that
\[
\frac{f^{(3)}(u_{c})}{V''(u_{c})}-\frac{V^{(4)}(u_{c}) f'(u_{c})}{V''(u_{c})^2} > 0.
\]
\end{theorem}

\section{Examples}

\subsection{Fourier scenario}
The mechanical interpretation of the Curie-Weiss model, that is associated to the F-type normal, has already been extensively discussed in a number of papers (see e.g. \cite{barraJSP}).
Let us briefly recall the main leading to the definition of such an analogy.
\begin{definition}
The Curie-Weiss Hamiltonian is defined by the Hamiltonian of the form
\be
\frac{1}{N}H_N(m_{N}) = -\frac{1}{2}m_N^2 + h m_{N},
\ee
\end{definition}
We are interested in an explicit expression of the free energy in terms of the order parameter. A number of methods has been proposed over the decades and are currently available (see e.g. \cite{barraJSP} for a recent review) to evaluate the free energy including a solution method based on a {\em mechanical analogy}.
\newline
Following the interpolation procedure introduced in~\cite{Gsumrules}, let us consider the interpolating free energy (or interpolating {\em action})
\be\label{euclide}
\alpha_N(t,x) = \frac1N \ln\sum_{\{ \sigma \}}^{2^N}\exp\left( -t \cdot \frac{N m_N^2}{2} + x \cdot m_N \right)=
\frac1N \ln\sum_{\{ \sigma \}}^{2^N}\exp\left(\bold{X}\cdot\bold{E} \right),
\ee
such that $\alpha(t=-\beta,x=0)=\lim_{N \to \infty}  \alpha_N(t=-\beta,x=0)$, i.e. it returns to the thermodynamical free energy in absence of external field.

Note that in the last term of eq.~(\ref{euclide}) we have introduced the two-vector space-time as $\bold{X}=(t,-x)$ and the two-vector energy-momentum as $\bold{E}/N=(\langle m_N^2\rangle/2, \langle m_N \rangle)$


\begin{figure}
\begin{center}
\includegraphics[width=3cm]{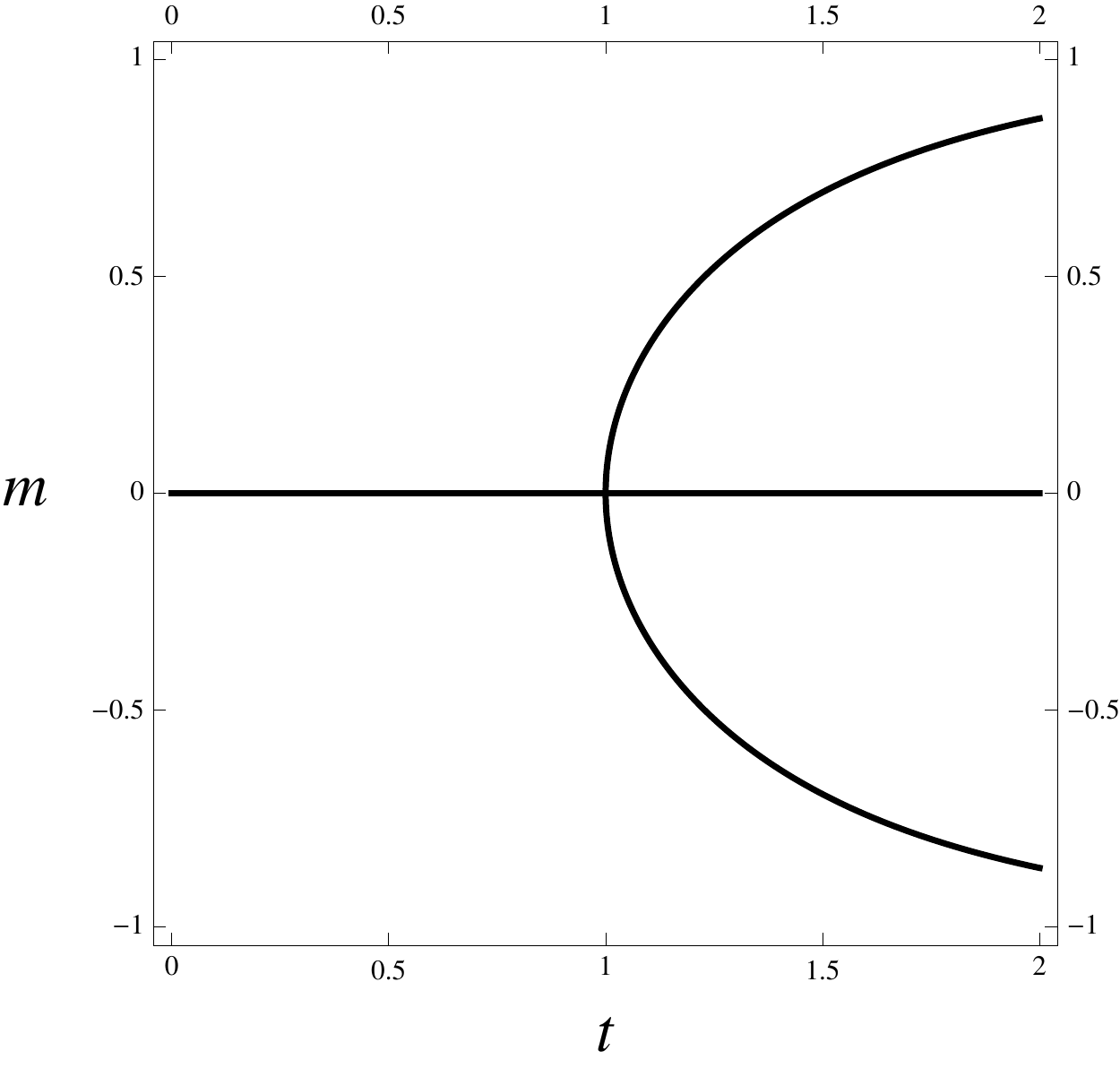}
\includegraphics[width=3cm]{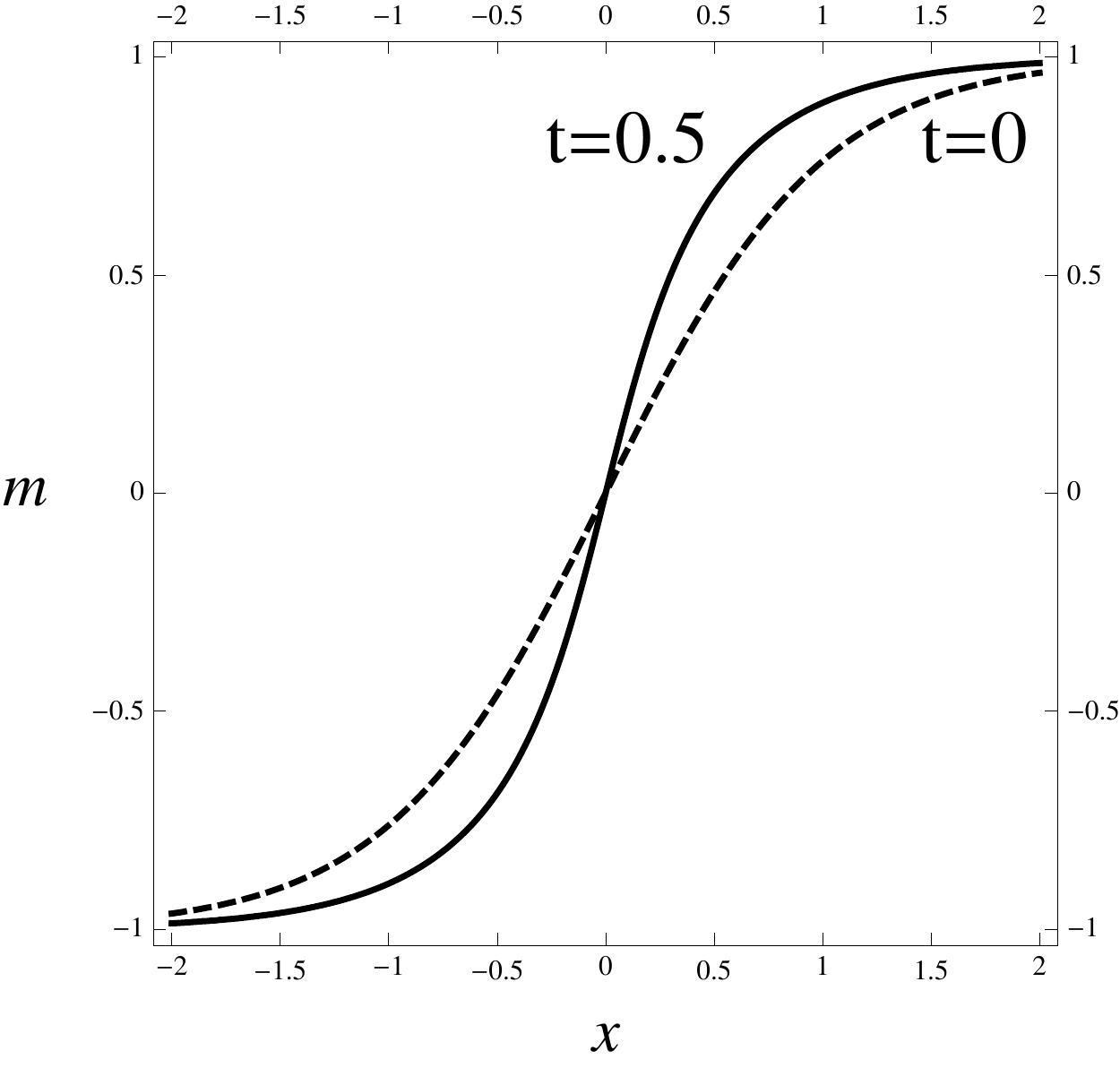}
\includegraphics[width=3cm]{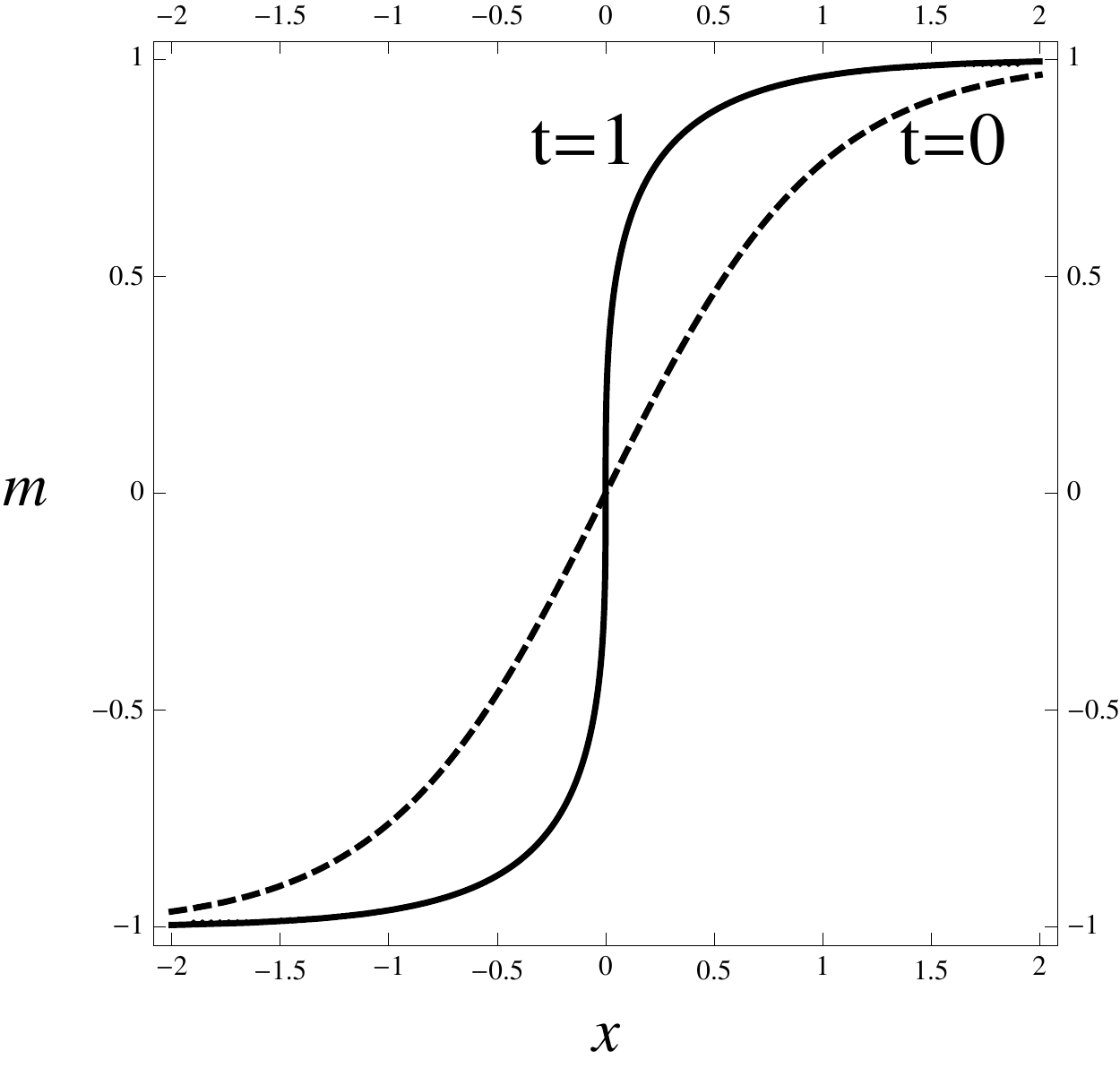}
\includegraphics[width=3cm]{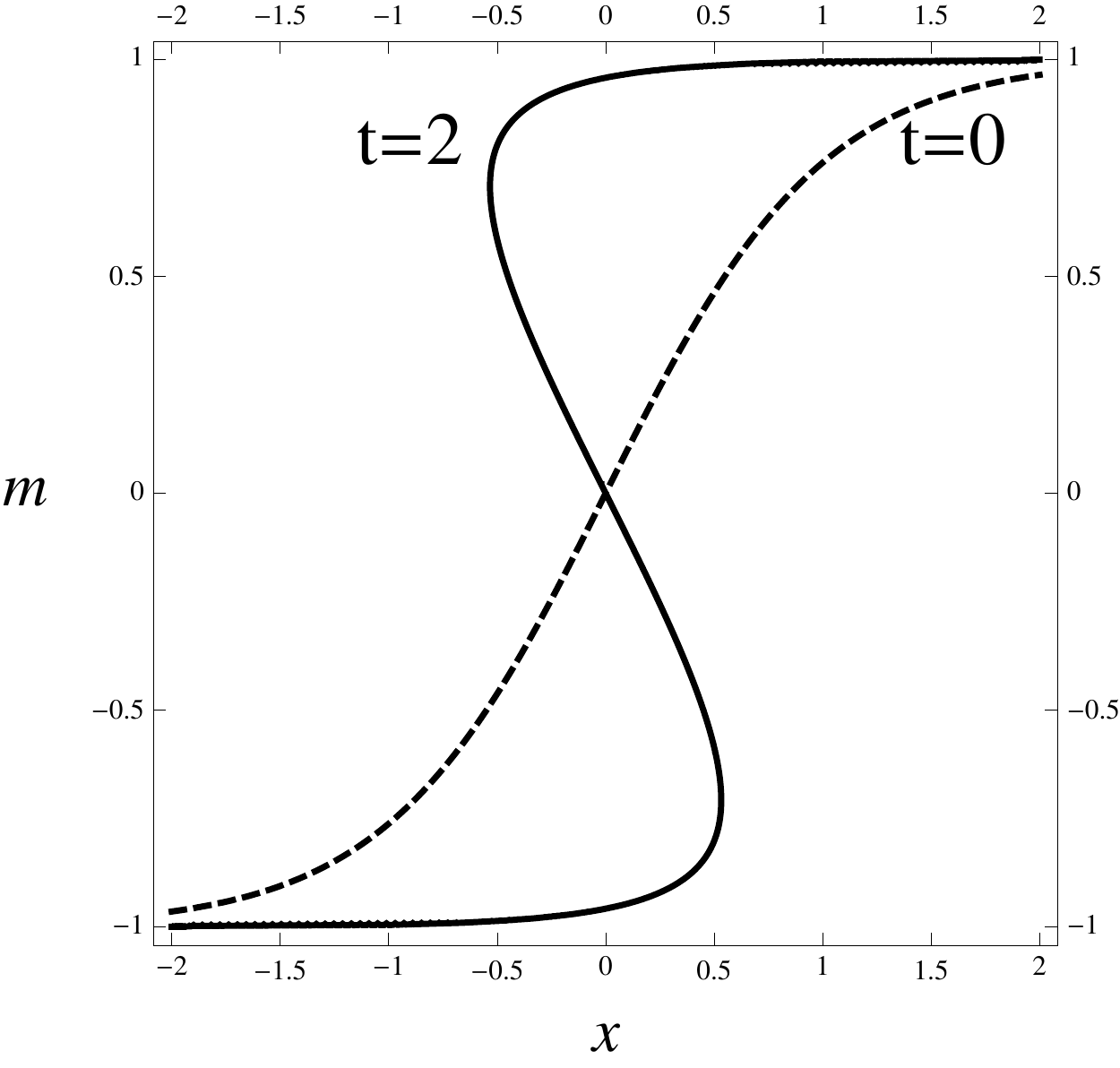}
\caption{\footnotesize{Analysis of the F-type. From left to right: Magnetization profile at $x=0$ versus $t$. Magnetization profile versus x at -respectively- $t=0.5 < t_c$, $t=1.0 =t_c$, $t=2 > t_c$. Beyond the gradient catastrophe that occurs at $t=1.0$ the solution exhibit a multivalued solution associated to metastable states of the system. The initial datum (at $t=0$) is reported too for visual comparison.}}
\label{fig:ft}
\end{center}
\end{figure}


\begin{theorem}[\cite{Gsumrules}]
The free energy~(\ref{euclide}) satisfies the following Hamilton-Jacobi type equation
\be\label{HJ-euclideo}
\frac{\partial \alpha_N(t,x)}{\partial t} + \frac12 \left( \frac{\partial \alpha_N(t,x)}{\partial x}\right)^2 - V_N(t,x)=0,
\ee
where $V_N(t,x) = N^{-1} \partial^{2}_{x} \alpha(x,t) =  \frac12 \left( \langle m^2_N \rangle - \langle m_N \rangle^2 \right)$.
\end{theorem}
\proof
By a direct calculation, it is straightforward to show that the expression~(\ref{euclide}) for the free energy solves the equation~(\ref{HJ-euclideo}).
\endproof

In the domain where the function $\alpha(x,t)$ is sufficiently smooth (i.e. smooth enough to have a unique maximizer in the variational problem of Theorem $1$), in the thermodynamic limit, we have
$$\lim_{N\to\infty}V_N(t,x) = \lim_{N\to\infty}\frac12 \left( \langle m^2_N \rangle - \langle m_N \rangle^2 \right)=0$$
and the corresponding free energy
\[
\alpha(t,x) = \ln 2  +\ln \cosh (x +m(t,x) t)-\frac{m(t,x)^{2}}{2} t
\]
is the solution to the Hamilton-Jacobi equation~
(\ref{HJ-euclideo})  with the initial datum
\be
\alpha(0,x) = \ln 2 + \ln \cosh x
\ee
that is obtained via a direct evaluation of the sum in~(\ref{euclide}) and where $m(t,x)$ is the unique maximizer in the variational problem defined by Theorem $1$.
In particular, at zero external field where the phase transition occurs we have \cite{genovese}
\be\label{alpha_CW}
\alpha(\beta) = \ln 2 + \ln\cosh(\beta m ) - \frac{1}{2}\beta m^2,
\ee
where we recall that $t =\beta$. \\

Remarkably, principles of thermodynamics (such as the free energy minimization) play here as the Maupertius minimim action principle and imply the extremization of this expression w.r.t.~the order parameter giving the celebrated self-consistency equation $\langle m \rangle = \tanh(\beta \langle m \rangle)$.
\newline
As it is well known, the self-consistency equation predicts a paramagnetic phase at $\beta < 1$, with $\langle m \rangle \equiv 0$ and a bifurcation at the critical noise level $\beta_c=1$, from which two branches of the magnetization (symmetric around zero) arise and the system undergoes a phase transition toward a ferromagnetic phase. As  Fig.~$1$ shows, the magnetization develops a gradient catastrophe at the origin $x=0$ where $m$ vanishes and at $t=1$. The critical values are obtained via the equations~(\ref{hodcrit}).

\subsection{Klein-Gordon scenario}
As discussed above the Curie-Weiss Hamiltonian is an F-type normal form~(\ref{Flead}) associated with the classical (Euclidean) kinetic energy.
Let us now focus on the $K$-type normal form~(\ref{Klead}) whose mechanical analogue can be viewed as a relativistic extension of the Curie-Weiss model.
\begin{definition}
The Hamiltonian of the K-type model  is defined as follows
\be
\frac{-H_N(m_{N})}{N}= \sqrt{1+m_N^2} + h m_{N}.
\ee
\end{definition}
Let us observe that introducing the variable $v$ (the relativistic speed) via $m=\gamma v$ with $\gamma = (1-v^{2})^{-1/2}$ we have $\sqrt{1+m^2}=(1-v^2)^{-1/2}$.
By a direct calculation, we can prove the following
\begin{theorem}
The interpolating action/free-energy reads as
\be
\alpha_N(t,x)=\frac1N \ln \sum_{\sigma}^{2^N}\exp\left(t \sqrt{1 + m_N^2} + x \cdot N m_N \right)=\frac1N \ln\sum_{\sigma}^{2^N}\exp\left(\bold{X}\cdot\bold{E} \right),
\ee
and obeys the following relativistic Hamilton-Jacobi equation
\begin{eqnarray}
\label{ciccio}
\left( \frac{\partial \alpha_N(t,x)}{ \partial t}\right)^2  - \left( \frac{ \partial \alpha_N(t,x)}{ \partial x}\right)^2 + V_N(t,x) = 1,\\
V_N(t,x)= \frac1N \left( (\partial^2_{tt}\alpha_N(t,x))-(\partial^2_{xx} \alpha_N(t,x))\right). \nonumber
\end{eqnarray}
\end{theorem}
\begin{figure}
\begin{center}
\includegraphics[width=3cm]{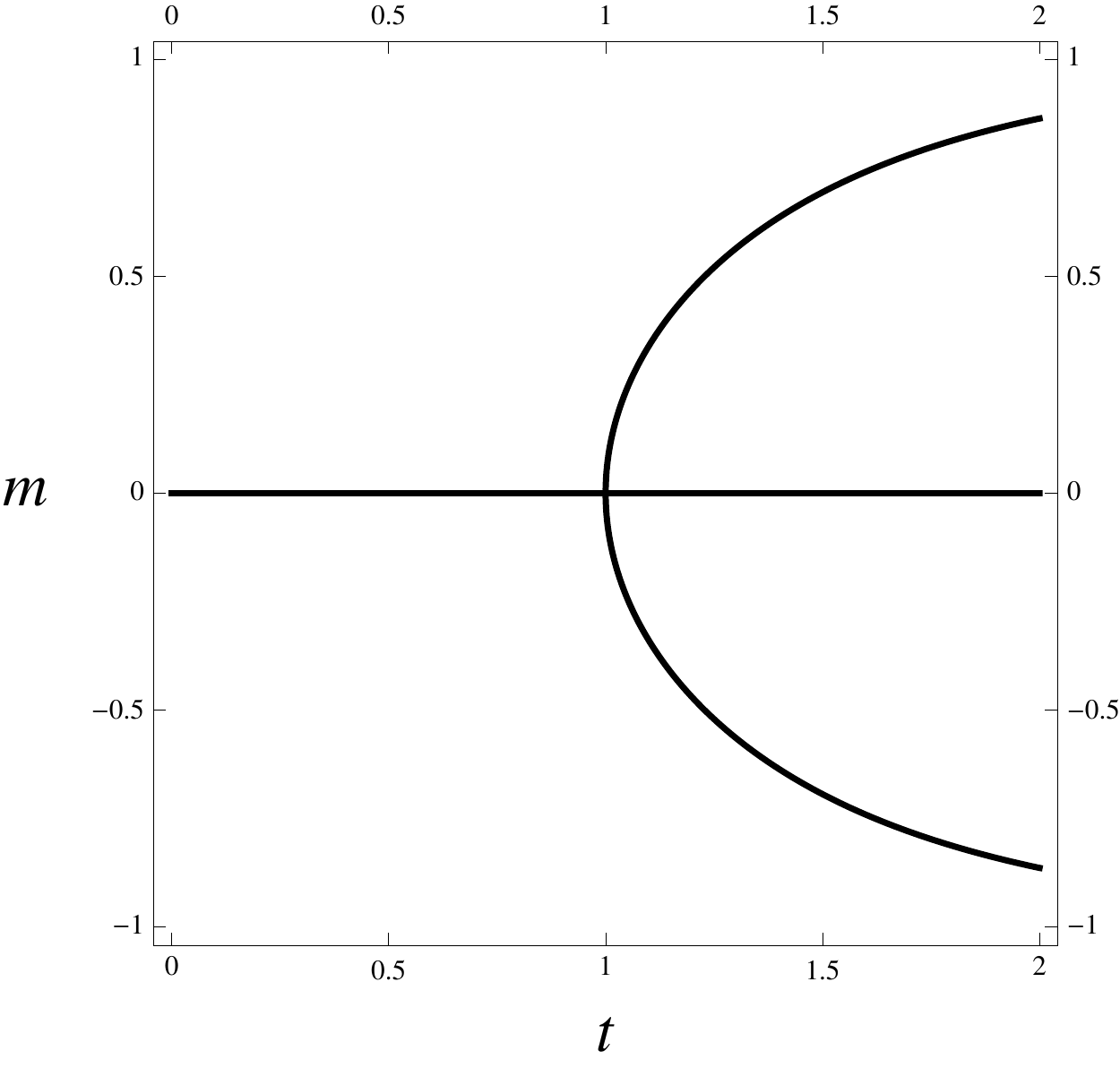}
\includegraphics[width=3cm]{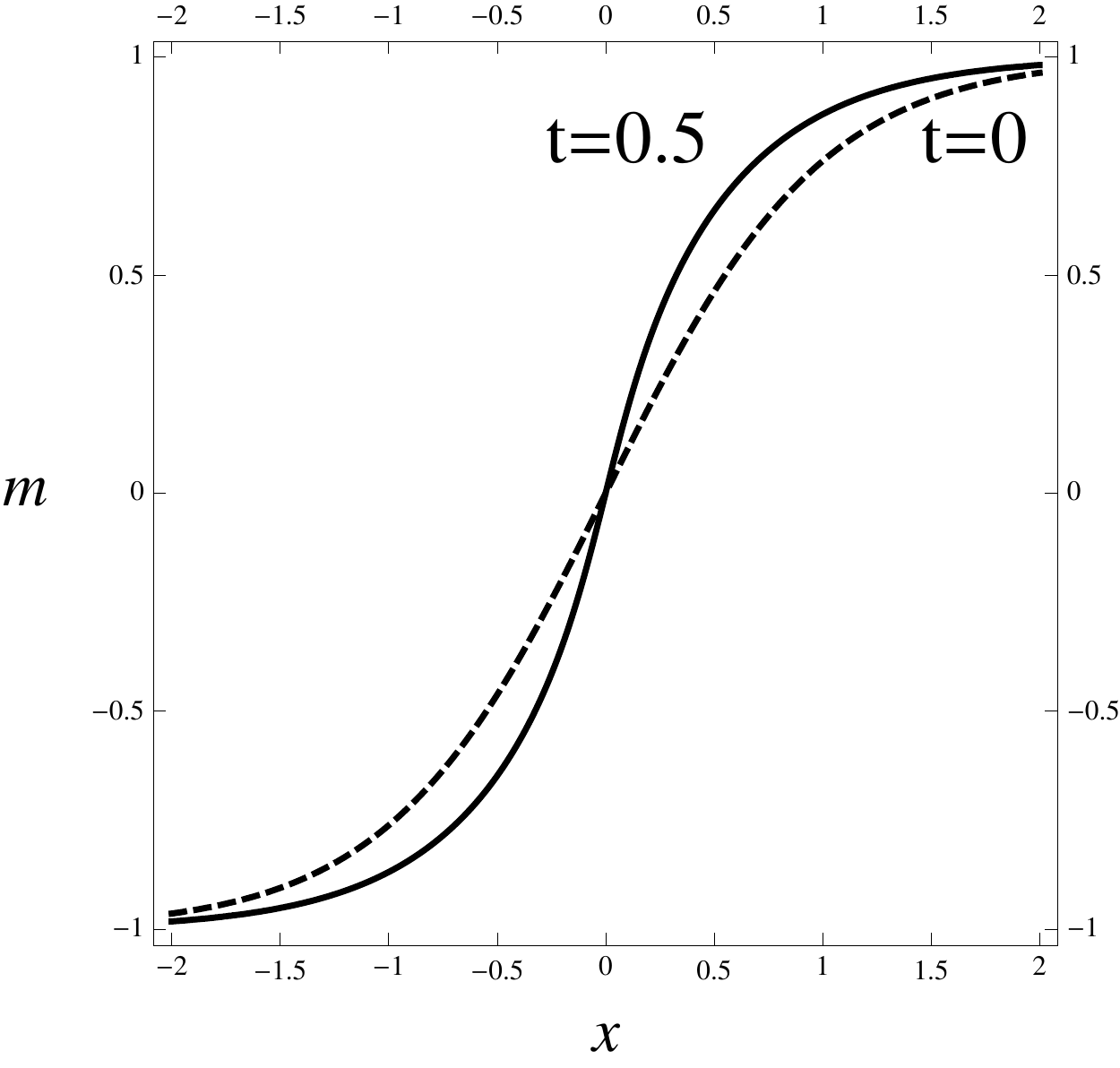}
\includegraphics[width=3cm]{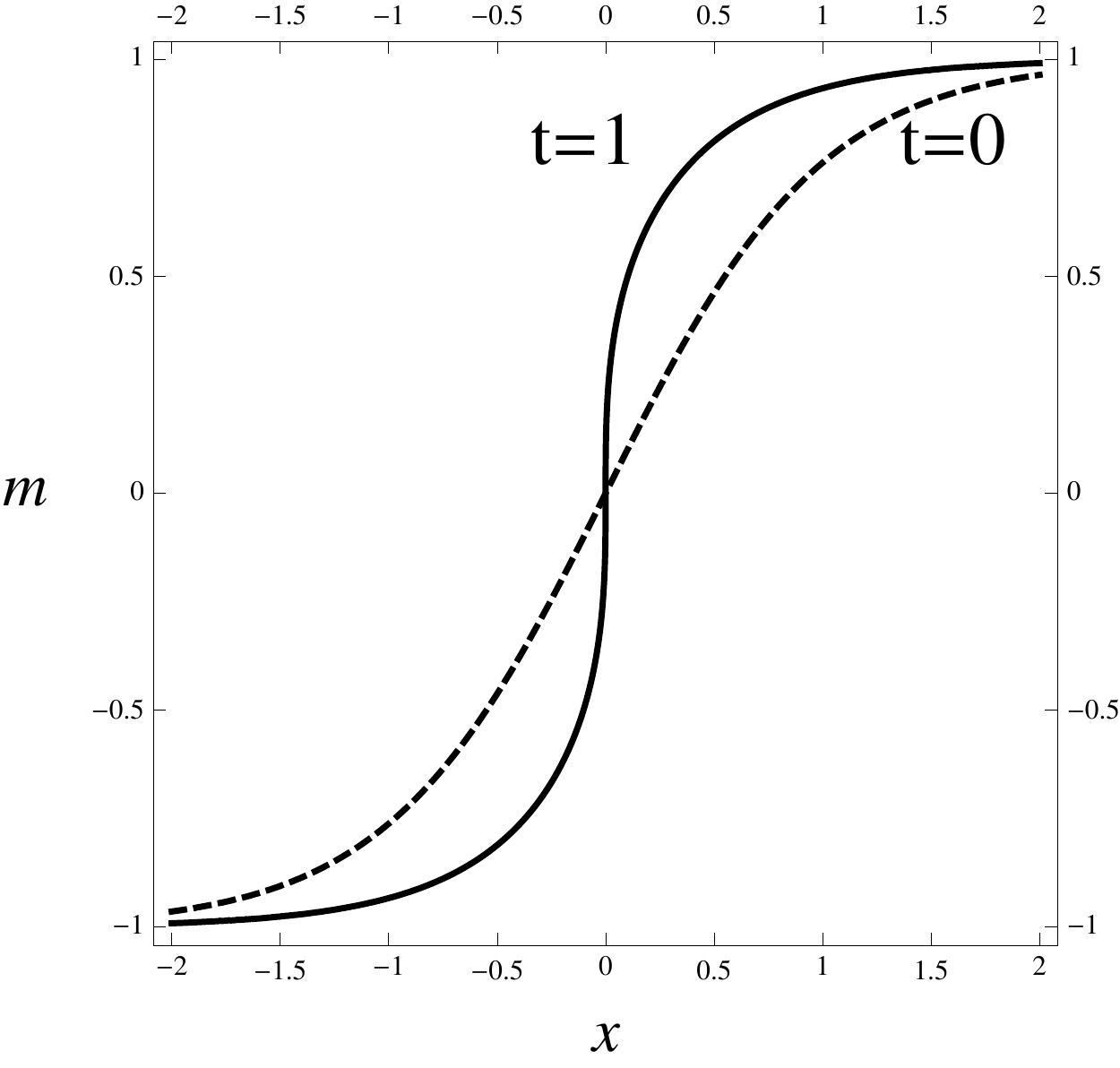}
\includegraphics[width=3cm]{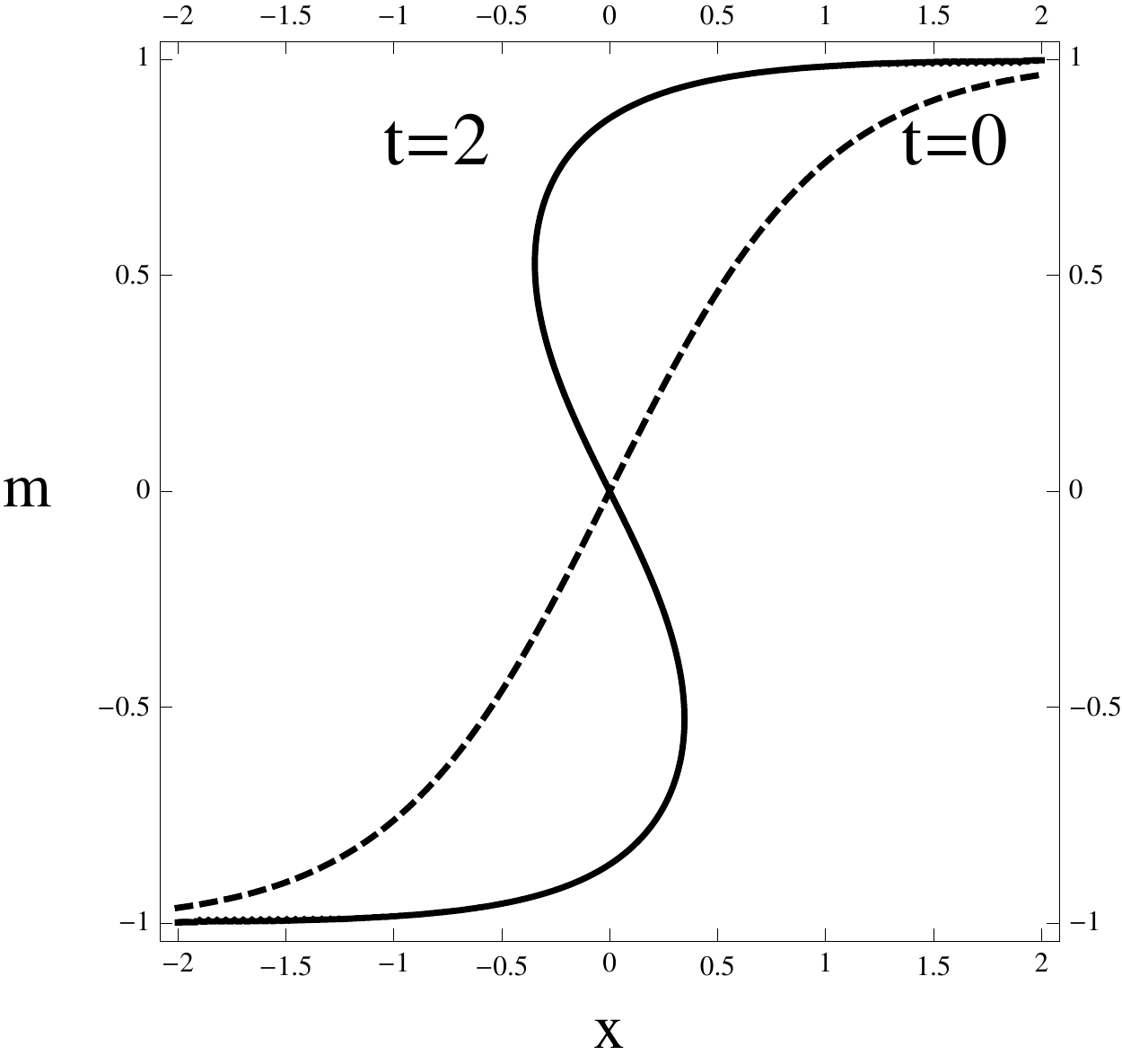}
\caption{\footnotesize{Analysis of the K-type. From left to right: Magnetization profile at $x=0$ versus $t$. Magnetization profile versus x at -respectively- $t=0.5 < t_c$, $t=1.0 =t_c$, $t=2 > t_c$. Similarly to the Curie-Weiss model, the magnetization profile breaks in the origin at $t=1$ and develops multivaluedness for $t>1$. The initial datum (at $t=0$) is reported too for visual comparison.}}
\label{fig:ft}
\end{center}
\end{figure}
%
%

Note that the potential is given, up to a scale factor $1/N$ , by the {\em D'Alambertian} of the action, that is a relativistic invariant,  and consequently, the left hand side of the Hamilton-Jacobi equation is also Lorentz-invariant.
\newline
As observed above, the thermodynamic free energy is obtained via the identification $t=\beta$ and $x=\beta h$.

\subsubsection{Generalized free energy by Minimum Action Principle}

Introducing the standard notation of covariant and contravariant vectors the  eq.~(\ref{ciccio}) reads as
\begin{equation}
\dfrac{\partial\alpha_{N}}{\partial x^{\mu}}\dfrac{\partial\alpha_{N}}{\partial x_{\mu}}+\dfrac{1}{N}\square\alpha_{N}=1 \label{eq:hj_rel_form_compl}
\end{equation}
and it can be interpreted as the Hamilton-Jacobi equation describing the motion of a relativistic particle in the potential  $V_{N}(t,x)= (\square\alpha_{N}(t,x))/N$.
\newline
We observe that, as in the Curie-Weiss case, the potential vanishes in the thermodynamic limit as long as the function $\alpha_N(t,x)$ is smooth.
\newline
Hence, in the thermodynamic limit, the equation (\ref{ciccio}) gives
\begin{equation}
\dfrac{\partial\alpha_{N}}{\partial x^{\mu}}\dfrac{\partial\alpha_{N}}{\partial x_{\mu}}= m_0 c^2 \equiv 1, \label{eq:hj_rel_form}
\end{equation}
which, from a field theory perspective, gives the semi-classical Klein-Gordon scenario~\cite{BD}.
\begin{remark}
In relativistic mechanics, the generalized momentum is defined as
\begin{center}
$P^{\mu}= \left (\dfrac{E}{c},\gamma mv \right),$
\end{center}
where $v$ is the classical velocity of the particle, $\gamma=\dfrac{1}{\sqrt{1-v^{2}}}$ is the Lorentz factor and $E=\gamma$ (we set the rest energy $m_0 c^{2}=1$) is the relativistic energy, hence, consistently with our findings, we have
\begin{equation}
\left (\dfrac{E}{c} \right )^{2}-(\gamma mv)^{2}=\dfrac{1}{1-v^{2}}-\dfrac{v^{2}}{1-v^{2}}=1.
\end{equation}

Moreover, observing that the covariant  gradient of the action is the contra-variant momentum (see e.g. \cite{gold})
\begin{center}
$\dfrac{\partial \alpha}{\partial x_{\mu}}=(\alpha_{t},-\alpha_{x})=P^{\mu}.$
\end{center}
we have the following identification between the statistical mechanical and relativistic  dynamical variables
\begin{equation}
P^{\mu}= (\gamma,\gamma v)=(\sqrt{1+m^{2}}, m)
\label{eq:statistical_quadri-momentum}.
\end{equation}
\end{remark}
\begin{remark}
Let us observe that the expansion of the energy in Taylor series around $m=0$, i.e.
\begin{center}
$E=\sqrt{1+m^{2}}=1-\dfrac{1}{2} m^{2}+O( m^{4})$
\end{center}
corresponds to the non-relativistic limit, where the leading order constant is identified with the rest energy (normalized as $m_0 c^{2} =1$) and the first order contribution is the Curie-Weiss potential associated to the Euclidean kinetic energy.
\end{remark}

\begin{Proposition}
The free energy of the K-type model at zero external field is
\begin{equation}
\alpha(\beta)=\ln 2+\ln\cosh\left(\dfrac{ m }{\sqrt{1+m^{2}}}\beta\right)+\dfrac{\beta}{\sqrt{1+m^{2}}}. \label{eq:free_energy}
\end{equation}
The associated self-consistency condition $\partial \alpha /\partial m = 0$ reads as
\be
\label{eq:Kself}
m = \tanh\left(\beta \dfrac{ m}{\sqrt{1+ m^{2}}}\right).
\ee
\end{Proposition}
\proof
Let us note that the equation~(\ref{eq:hj_rel_form}) describes the  free motion of a relativistic particle and can be readily integrated. Observing that the relativistic Lagrangian $\mathcal{L}=-\gamma^{-1}$ is preserved along the characteristics $x+vt$, then the action is computed as follows
\begin{eqnarray}\label{eq:free_energy_interpolated}
\alpha(t,x)&=&\alpha(0,x)+\int\limits_{0}^{t}\dfrac{dt'}{\gamma}= \ln 2+\ln\cosh(-x)+\dfrac{t}{\gamma}\\ \nonumber
&=& \ln 2 + \ln\cosh(vt-x)+\dfrac{t}{\gamma} =\ln 2+\ln\cosh(\dfrac{ m}{\sqrt{1+m^{2}}}t-x)+\dfrac{t}{\sqrt{1+m^{2}}}.
\end{eqnarray}
Evaluating $\alpha (\beta,0)$ one obtains the solution~(\ref{eq:free_energy}).
\endproof
\begin{remark}
Let us observe that, free energy and self-consistency equation for the Curie-Weiss model are readily recovered from the Taylor expansion around $m=0$ of the equations~(\ref{eq:free_energy}) and ~(\ref{eq:Kself}) respectively.
\end{remark}

\subsection{Poisson scenario}
We finally discuss the case of the elliptic dispersion curve.
\begin{definition}
The Hamiltonian of the P-type model is defined as follows
\be
\frac{-H_N(m_{N})}{N}= -\sqrt{1-m_{N}^2} + h m_{N}.
\ee
\end{definition}
%
As discussed above, the partition function is obtained as a solution to the Poisson equation~(\ref{Ptype}). Moreover the free energy $\alpha = -\nu \log Z$ in the thermodynamic limit satisfies the equation~(\ref{Plead}) and it is given according to the following
\begin{theorem}
Fixing $h=0$, the free energy of the generalized ferromagnetic P-type model coupled is
\begin{equation}
\label{libera}
\alpha(\beta)=\ln 2+\ln\cosh \left (\beta\dfrac{m}{\sqrt{1- m^{2}}} \right)-\dfrac{\beta}{\sqrt{1- m^{2}}}.
\end{equation}
Moreover, the self-consistency equation reads as follows
\begin{equation}
 m =\tanh\left(\beta\dfrac{ m}{\sqrt{1- m ^{2}}}\right).
 \label{selfcons2}
\end{equation}
\end{theorem}


\begin{figure}
\begin{center}
\includegraphics[width=3cm]{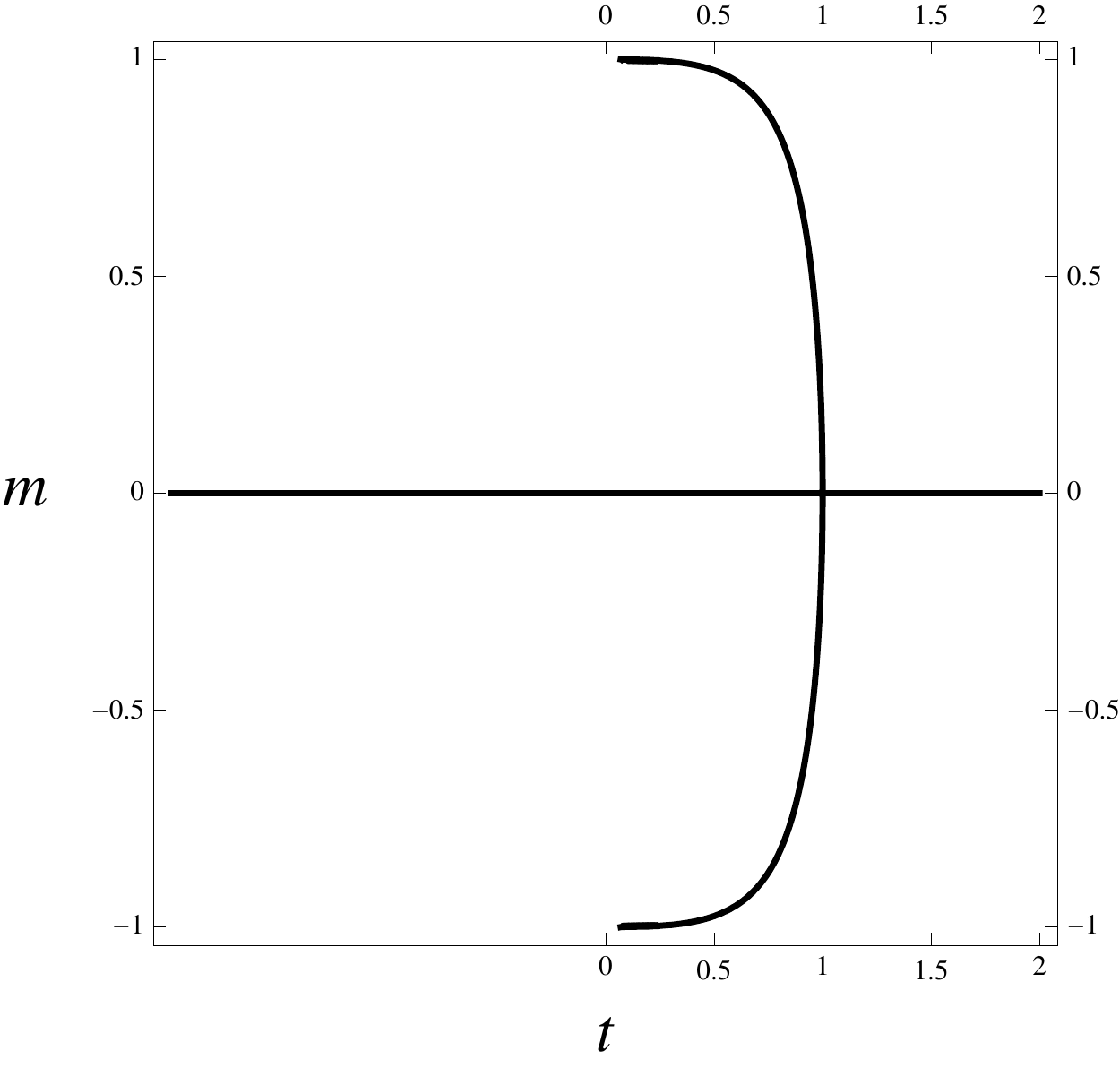}
\includegraphics[width=3cm]{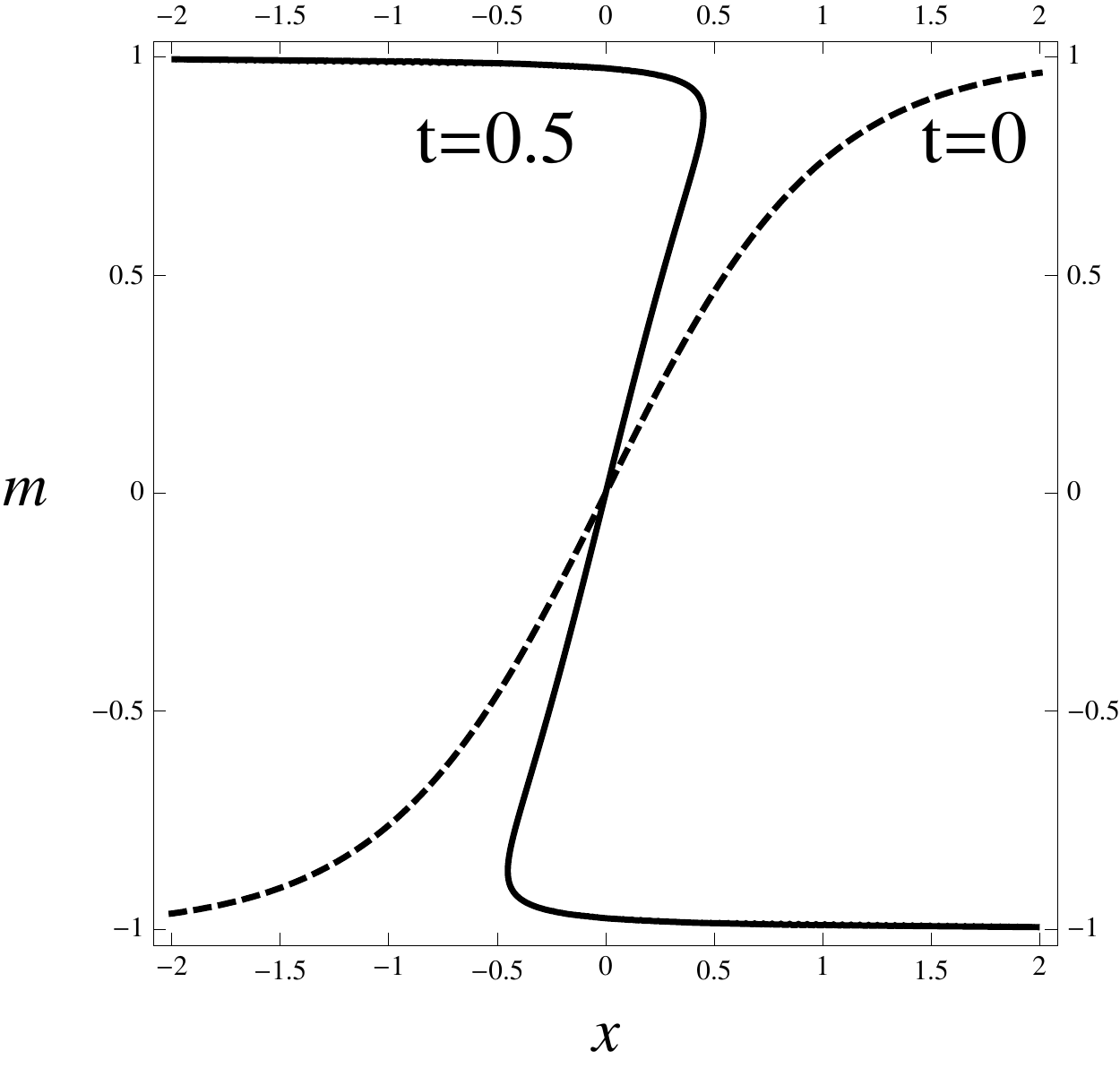}
\includegraphics[width=3cm]{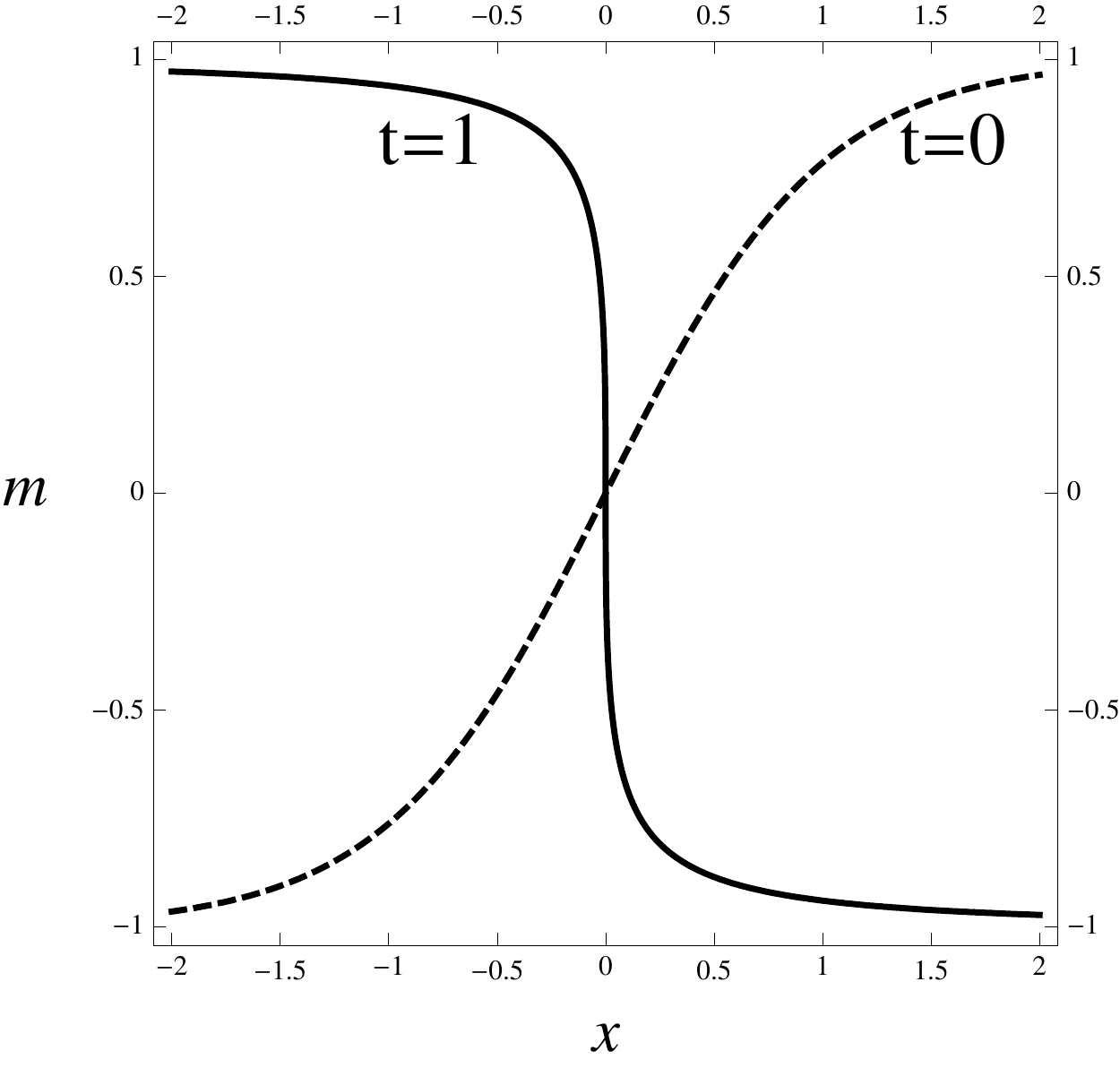}
\includegraphics[width=3cm]{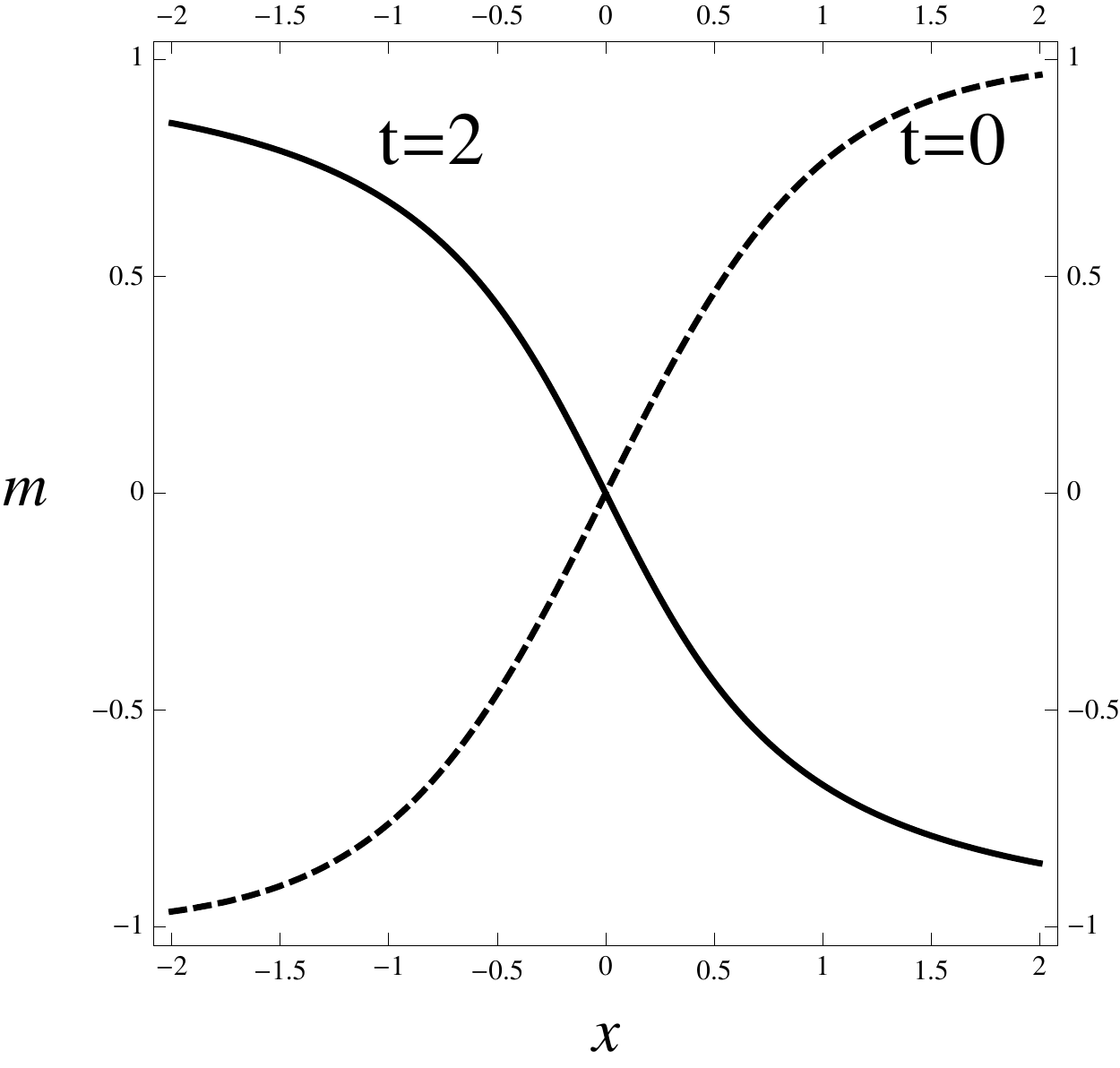}
\caption{\footnotesize{Analysis of the P-type. From left to right: Magnetization profile at $x=0$ versus $t$. Magnetization profile versus x at  $t=0.5 < t_c$, $t=1.0 =t_c$, $t=2 > t_c$ respectively. The initial datum (at $t=0$) is also reported for a visual comparison. Note that in the high noise region, in addition to the (stable) solution say $m=0$, two additional (instable) extremal points (maxima) for the free energy appear as consequence of the infinite ferromagnetic contributions.}}
\label{fig:ft}
\end{center}
\end{figure}
\begin{remark}
Similarly to the K-type case, free energy and self-consistency equation for the Curie-Weiss model are readily recovered from the Taylor expansion around $m=0$ of the equations~(\ref{libera}) and ~(\ref{selfcons2}) respectively.
\end{remark}
As shown in figure~(\ref{fig:ft}), due to the ill-posedness of the initial value problem, the solutions does not evolve continuously from the initial datum producing a multi-valued solution due to occurrence of additional two instable extremal points for the free energy as consequence of the infinite ferromagnetic contributions.


\section{Conclusions}
In this paper we have discussed in detail a {\em formal analogy} between
the thermodynamic evolution of mean-field spin systems and one dimensional Hamiltonian
systems.

We focussed our attention on the class of spin models associated to an algebraic dispersion curve that contains the celebrated Curie-Weiss model as a particular case. The partition function for a finite number $N$ of particles plays the role the {\it quantum} wave function and obeys a linear PDE. The thermodynamic limit is obtained
via the standard WKB analysis, where the Hamilton Principal Function is identified with the free energy of the thermodynamic system. The Hamilton-Jacobi equation can be treated via standard techniques and it is showed that the magnetization is a solution to a Riemann-Hopf type equation. Hence, the model is completely integrable and solvable by the characteristics method.

Within this framework, thermodynamic phase transitions are associated to the occurrence of caustics in the semiclassical approximation. In particular, the critical point is identified with the point of gradient catastrophe where the magnetization satisfies the Riemann-Hopf equation.
\newline
All these features are discussed in detail for the class of models associated to a second order dispersion curve. The reduction of the dispersion curve to the canonical form leads to three family of models associated with the conics: F-type - parabolic, K-type-hyperbolic and P-type-elliptic.
F-type models are associated to the semiclassical dynamics of a non-relativistic particle. Such models are reduced to the Curie-Weiss model that has been extensively  studied in the literature~(see e.g. \cite{barbone, genovese}). K-type models give a class of infinitely many $p-spin$ contributions (namely higher order interactions in the Hamiltonian, e.g., from $m^2$, to $m^4, m^6, ..., m^p$) to the interaction and the thermodynamic limit is associated to the semiclassical limit of a relativistic particle. P-type models describe infinitely many ferromagnetic $p-spin$ contributions to the interaction associated to an elliptic dynamics. In particular we observe that due to the ill-posedness of the initial value problem, ferromagnetic contributions sum up to produce two meta-stable states (local maxima of the free energy) in the ergodic region.
\newline
We observe that both K-type and P-type extensions of the Curie-Weiss model can be viewed as ``relativistic" extensions of the Curie-Weiss model as the speed remains bounded, although only the K-type is associated to a Lorentz invariant Hamiltonian system.

\section*{Acknowledgments}
The authors are pleased to thank Paolo Lorenzoni for useful references and discussions.
AB is grateful to GNFM-INdAM Progetti Giovani 2014 grant on {\it Calcolo parallelo molecolare}, for financial support.
FG is grateful to INFN Sezione di Roma for financial support.
AM is grateful to GNFM-INDAM Progetti Giovani 2014 grant on {\it Aspetti geometrici e analitici dei sistemi integrabili},  the London Mathematical Society Visitors Grant (Scheme 2) Ref.No.~21226 and Northumbria University starting grant for financial support.




\begin{thebibliography}{9}









\bibitem{ALM}
A. Arsie, P. Lorenzoni, A. Moro, \emph{Integrable viscous conservation laws},  {\tt arXiv:1301.0950} (2013).



\bibitem{barraJSP} A. Barra, {\em The mean field Ising model trough interpolating techniques}, J. Stat. Phys. \textbf{132}(5), 787-809 (2008).



\bibitem{aldo} A. Barra, A. Di Biasio, F. Guerra, {\em Replica symmetry breaking in mean-field spin glasses through the Hamilton-Jacobi technique}, J. Stat. Mech. \textbf{09}, 09006, (2010).


\bibitem{Fourier} A. Barra, G. Del Ferraro, D. Tantari, {\em Mean field spin glasses treated with PDE techniques}, E. Phys. J. B \textbf{86}, 332, (2013).
	
	



\bibitem{BD} J.D. Bjorken, S.D. Drell, {\em Relativistic quantum mechanics}, New York: McGraw-Hill (1964).

\bibitem{bogo1} N. Bogolyubov,  et al. {\em Some classes of exactly soluble models of problems in Quantum Statistical Mechanics: the method of the approximating Hamiltonian}, Russian Mathematical Surveys \textbf{39},(6): 1-50, (1984).

\bibitem{bogo2} J.G. Brankov, A.S. Shumovsky, V.A. Zagrebnov, {\em On model spin Hamiltonians including long-range ferromagnetic interaction}, Physica \textbf{78},(1):183, (1974).

\bibitem{bogo3} J. G. Brankov and V. A. Zagrebnov, {\em On the description of the phase transition in the
Husimi-Temperley model}, J. Phys. A: Math. Gen. \textbf{16}, (1983) 2217-2224.

\bibitem{barbone} P. Choquard, J. Wagner, {\em On the ''Mean Field'' Interpretation of Burgers' Equation}, J. Stat. Phys. \textbf{116}:843-853, (2004).


\bibitem{moro2} G. De Nittis, A. Moro, {\em Thermodynamic phase transitions and shock singularities}, Proc. R. Soc. A \textbf{468}, 701-719 (2012).


\bibitem{DE} B. Dubrovin, M. Elaeva, \emph{On critical behaviour in nonlinear evolutionary PDEs
with small viscosity},  Russian Journal of Mathematical Physics, vol. 19, p. 13-22.



\bibitem{moro3} G. De Nittis, P. Lorenzoni, A. Moro, {\em Integrable multi-phase thermodynamic systems and Tsallis' composition rule}, J. of Phys. Confer. Series \textbf{482}(1):012009, (2014).


\bibitem{gardner} E. Gardner, {\em Spin glasses with P-spin interactions}, Nucl. Phys. B \textbf{257}: 747-765 (1985).

\bibitem{genovese} G. Genovese, A. Barra, {\em A mechanical approach to mean field spin models}, J. Math. Phys. \textbf{50}(5), 053303 (2009).


\bibitem{gold} H. Goldstein, {\em Classical Mechanics}, Pearson Education, Edimburgh (2014).


\bibitem{Gsumrules} F. Guerra, {\em Sum rules for the free energy in the mean field spin glass model}, Fields Institute Communications \textbf{30}, Amer. Math. Soc. (2001).


\bibitem{hightempGT} F. Guerra, F.L. Toninelli, {\em The thermodynamic limit in mean field spin glass models}, Comm. Math. Phys. \textbf{230}(1), 71-79 (2002).


\bibitem{I} A.M. Il'in,  \emph{Matching of Asymptotic Expansions of Solutions of Boundary Value Problems},
AMS Translations of Mathematical Monographs, Vol. 102, (1992).







\bibitem{moro1} A. Moro, {\em Shock dynamics of phase diagrams}, Annals Phys. \textbf{343}, 49-60 (2014).





\bibitem{ruelle} D. Ruelle, {\em Statistical mechanics: Rigorous results}, World Scientific 1999.
\bibitem{shannon} S. Shannon, {\em Thermodynamic Limit for the Mallows Model on $ S_n$}, arXiv preprint arXiv:0904.0696 (2009).




\bibitem{Whi} G.B. Whitham, Linear and Nonlinear Waves,  1974, Wiley, New York.

\end{thebibliography}
\end{document}